\documentclass[
 amsmath,amssymb,
 aps, physrev,
]{revtex4-2}

\usepackage{graphicx}
\usepackage{dcolumn}
\usepackage{bm}

\begin{document}

\preprint{}

\title{\textbf{No missing flare in OJ~287} 
}%

\author{Mauri J. Valtonen}
 \altaffiliation{}
\affiliation{FINCA, University of Turku, FI-20014 Turku, Finland}
\affiliation{Tuorla Observatory, Department of Physics and Astronomy, University of Turku, FI-20014 Turku, Finland}
\affiliation{Institute of Astronomy, University of Cambridge, Madingley Road, Cambridge CB3 0HA, UK \footnote{visitor}}
 \email{mvaltonen2001@yahoo.com}



\date{\today}

\begin{abstract}
The quasar OJ~287 has shown large flares since 1888, following a pattern that arises in a supermassive black hole binary when the secondary hits the accretion disk of the primary, and releases a hot bubble of gas at every disk crossing. A complete mathematical solution of the flare sequence produced a list of future flares, the latest happening in the summer of 2022.  Here I look into the origin of the idea that the lack of seeing the 2022 flare is a theoretical problem. During the summer OJ~287 cannot be observed by ground-based optical telescopes. In a paper published in 2021, ahead of the 2022 observing campaign, this was clearly stated. The often repeated claim that there is a "missing flare problem", is a misunderstanding, as no detection was possible with the current instrumentation. 
\begin{description}
\item[
To appear in RNAAS.]
\item[
keywords: Blazars; Active Galactic Nuclei; BL Lacertae objects: individual (OJ 287)]
\end{description}
\end{abstract}

\maketitle

\section{Introduction}

OJ~287 is a blazar/quasar at the redshift 0.306. It is one of the brightest quasars and its images can be found in old photographic plates since 1888. More recently, it has been intensively observed due its periodic properties. A recent description of the light curve and its explanation using a binary black hole model is found in \citet{val24}.
   
The binary model assumes an unequal mass black hole binary with an accretion disk. Every impact of the secondary produces a bubble of high temperature gas on both sides of the disk. The bubble initially radiates as a synchrotron source of flat spectral index, due to recent shock acceleration of relativistic electrons. With subsequent expansion it is seen as a large bremstrahlung flare when the bubble becomes optically thin.
   
Since the discovery of the periodic nature of OJ~287 in 1982, large flares fitting this description have been observed in 1983, 1984, 1994, 1995, 2005, 2007, 2015 and 2019. It should be noted that the large flares come in doubles and in occasional triples. The first triple was observed in 1957, 1959 and 1964, while the next triple is 2015, 2019 and 2022.

Using the timings of ten bright flares one finds a unique mathematical orbit solution with nine parameters. The model uses Post-Newtonian General Relativity to the order of 4.5, a theory of expanding shock-heated bubbles, disk level calculations at the times of impacts, and spin-orbit coupling. The model is over-determined since it also explains the timing of eight other flares. There are no free parameters that could be used to improve the model.

\section{The 2022 flare}

The 2022 flare was expected to start at 2022.548. Because there was an inevitable gap in observations from 2022.52 to 2022.66, \citet{val21} conclude that "The first big flare during 2022 is not expected to be visible from any Earth-based facility".

\begin{figure}
  \centering
   \includegraphics[width=0.8\textwidth]{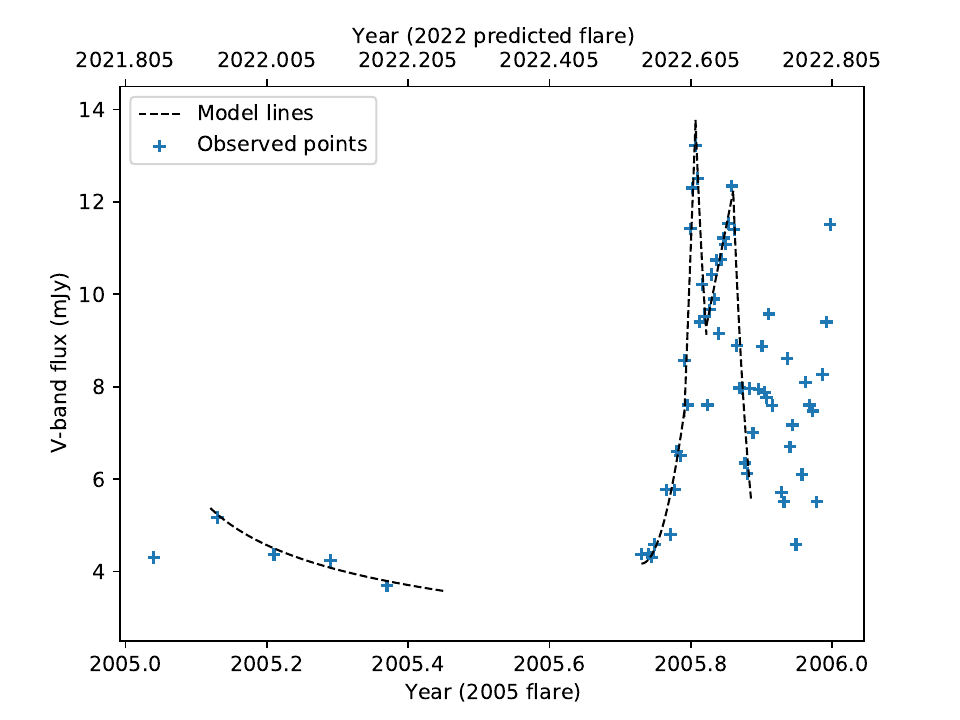}
   \caption{The optical light curve of OJ~287 during 2005 in V-band mJy units, in averaged data points. The dashed-line curve gives the general outline. If the disk levels and the impact distances were the same in 2005 and in 2022, one can draw an upper scale for the expected light curve in 2022 \citep{val21}.}
   \label{fig:2005_lc_fit}
\end{figure}

We know when the secondary crosses the average mid-plane of the accretion disk. However, until the position of the actual disk plane is calculated, we don't know the time of the impact. For impacts prior to 2022 the disk levels were calculated in \citet {val07}. However, back in 2007, the year 2022 seemed so far away that I did not include it.

The range of disk levels is from -75 AU to 265 AU, with the positive sign meaning that the secondary pulls the disk toward itself. In order that 2022 flare would be seen, the disk has to be below -500 AU, meaning that the approaching secondary would \textit{push away} the disk.

This calculation was reported in the second draft of \citet{val23} (posted as \citet{val22} in arXiv), but it was removed from the next draft and the final paper, since by then the disk level calculation was done. The disk was found at its mean level, and the impact point a little closer to the center than in 2005. This made the bubble expansion also faster. Combining this new theoretical knowledge with the detection of the early synchrotron flare in March 2022 (reported in the first draft in June 2022) it became clear that the big flare had happened 10 days earlier than was initially expected. Even the tail end of the flare was out of reach of the ground-based observations. Theoretically, there was no chance of seeing any trace of the big flare, and observations agreed.

In Figure \ref{fig:2005_lc_fit} we outline the 2005 flux evolution by dashed lines. The timescale for the expected flux evolution in 2022 is displayed in the upper scale, assuming that the 2022 and 2005 disk levels and light curves were the same. This graph was drawn before the 2022 disk level calculation: we have to slide the upper x-axis forward by 0.027 units to make the flare happen 10 days earlier. Now the summer gap in observations covers rather exactly the part of the 2005 light curve which is marked by the dashed line on the right hand side of the figure. Disregarding this section, the 2005 and 2022 light curves match well.

So what was missing then? Initially, I recommended observers carry out the campaign past the end of summer, since the disk level had not yet been calculated. This was misunderstood as a suggestion that the big flare should happen after the summer, and this showed up in the second draft \citep{val22}. The last sentence of the abstract, 'Observing the next impact flare of OJ~287 in October 2022 will substantiate the theory of disk impacts in binary black hole systems', was so poorly formulated, that it caused some people to expect something special in October. It was contrary to the main text which made clear that a flare in October would have required an unreasonable and previously unseen disk level. This sentence was meant to refer to the theory of disk impacts, not to their timing. The latter would have been awkward, and with later knowledge, contrary to the theory, if the flare had happened in October. The next draft, soon thereafter, corrected these misunderstandings.

\citet{kom22} quote \citep{val22}, rather than the next draft with corrections. Other authors, e.g. \citet{par24} have followed the practice and so the misunderstandings have snowballed. 

\begin{acknowledgments}
 This work has been supported by the Finnish Society of Sciences and Letters as well as the Institute of Astronomy at University of Cambridge.
\end{acknowledgments}


\bibliographystyle{aasjournal}





\end{document}